\documentclass[aps,preprint,epsfig,rotate]{revtex4}
\usepackage{graphicx}
\usepackage{bm}
\usepackage{epsfig}

\input epsf
\begin{document}
\title{Atomic analysis of the $(n;t)-$reaction of the helium-3 atoms with
       slow neutrons}

 \author{Alexei M. Frolov}
 \email[E--mail address: ]{afrolov@uwo.ca}

 \author{David M. Wardlaw}
 \email[E--mail address: ]{dwardlaw@uwo.ca}

\affiliation{Department of Chemistry\\
 University of Western Ontario, London, Ontario N6H 5B7, Canada}

\date{\today}

\begin{abstract}

Probabilities of formation of various hydrogenic species during the
exothermic nuclear $(n,{}^3$He$;t,p)-$reaction of atomic helium-3 with slow
neutrons are determined. In particular, we have found that the probability
to form the tritium atom ${}^3$H in its ground state is $\approx$ 55.19287
\%, while the analogous probability to form the protium atom ${}^1$H in its
ground state is $\approx$ 1.02363 \%. Analogous probabilities of formation
of the negatively charged hydrogen ions, i.e. the ${}^3$H$^{-}$ and
${}^1$H$^{-}$ ions, in the nuclear $(n,{}^3$He$; t,p)-$reaction with slow
neutrons, are $\approx$ 7.8680 \% and $\approx$ 0.06583 \%, respectively. We
also consider bremsstrahlung from fast fission-type reactions in atomic
systems. The spectrum of emitted radiation is analyzed.

\end{abstract}
\maketitle
\newpage

\section{Introduction}

The reaction of the ${}^{3}$He nuclei with slow neutrons is written in the
form \cite{Kik}
\begin{eqnarray}
 {}^3{\rm He} + n = p + t + 0.764 \; \; MeV \label{e1}
\end{eqnarray}
where the notations $p$ and $t$ stand for the protium ${}^{1}$H and tritium
${}^{3}$H nuclei, respectively. The cross-section $\sigma$ of this nuclear
reaction is extremely large for slow neutrons \cite{WW} ($\sigma_{max}
\approx 5330 \cdot 10^{-24}$ $cm^2$ or 5330 barn, for short). The
velocities of the two nuclear fragments formed in the reaction,
Eq.(\ref{e1}), are $\approx$ 1.59632 $a.u.$ (tritium) and $\approx$ 4.78897
$a.u.$ (protium). All values in this study are given in atomic units, where
$\hbar = 1, m_e = 1, e = 1$. The unit of atomic velocity is $v_e = \alpha c
\approx \frac{c}{137} \approx 2.1882661 \cdot 10^{8}$ $cm \cdot sec^{-1}$,
where $c$ is the speed of light and $\alpha$ is the fine structure constant.
The atomic velocity $v_e$ is the velocity of the $1s-$electron in the
hydrogen atom with the infinitely heavy nucleus ${}^{\infty}$H.

In our earlier work \cite{FrWa09} we have determined the probabilities to
detect various final (bound) states in hydrogen atoms which arise during the
exothermic nuclear $(n,{}^3$He$;t,p)-$reactions of the one-electron helium-3
ion (${}^3$He$^{+}$) with slow neutrons. In many actual experiments,
however, the nuclear $(n,{}^3$He; $t,p)-$reaction involves the two-electron
helium-3 atom rather the bare helium-3 nucleus. Analogous nuclear reactions
of the ${}^6$Li, ${}^{10}$B, ${}^{14}$N nuclei (see, e.g., \cite{Kik},
\cite{FrWa09}) with slow neutrons also involve few-electron atoms and ions,
rather than the bare nuclei and/or one-electron ions. Consequently the
products of such reactions may also contain a few bound electrons. Our aim
below is to evaluate the probabilities of formation of such few-electron
species.

In order to determine the probabilities of formation of few-electron atoms
and ions during nuclear reactions one needs to solve a number of problems.
Some of these problems are related to the analysis of the basic principles
of perturbation theory in quantum mechanics \cite{LLQ}. In particular, the
theory of sudden approximations \cite{Mig1}, \cite{Mig2} plays a central
role in our analysis below. Based on the sudden approximation we have
developed a transparent and reliable method which is used to determine the
final state probabilities for each possible system/state which can arise
after the nuclear reaction, Eq.(\ref{e1}). On the other hand, it is not
clear what procedure can be used for actual computations of the final state
probabilities. In \cite{FrWa09} we have described a general method based on
the use of $N_{f}-$electron density matrixes for the $N_{i}$-electron atomic
systems (here $N_{f} \le N_{i}$). This is a rigorous method, but in real
applications it produces a very complex procedure. In many cases this
procedure leads to the loss of numerical accuracy, e.g., in those cases when
one uses highly accurate variational wave functions constructed for
few-electron atomic systems. Another group of complications is related to
the electron-electron correlation which cannot be ignored in the incident
${}^{3}$He atom and final H$^{-}$ ion.

In this study we have re-considered all mentioned problems and developed an
approach which can be used in actual calculations of the final probabilities
in arbitrary atomic systems undergoing nuclear reactions. All details of our
analysis will be published elsewhere \cite{FroRui}. Here we want to apply
this method to obtain the actual data for the nuclear $(n,{}^3$He$;
t,p)-$reaction in the two-electron helium-3 atom. A successful test of our
method for the ${}^3$He atom with the exothermic nuclear $(n; t)-$reaction,
Eq.(\ref{e1}), is extremely important. If our method works in this case,
then it can be applied to other similar reactions in few-electron atomic
systems. Another goal of this study is to analyze radiation emitted during
fast fission-type reactions in atomic systems. In particular, we consider
the spectrum of bremsstrahlung emitted in the nuclear reaction,
Eq.(\ref{e1}), of atomic helium-3 with slow neutrons.

\section{Formation of the tritium/protium atoms and ions}

Note that all current evaluations of the final state probabilities in
various atomic systems with nuclear reactions are based on the sudden
approximation \cite{Mig1}, \cite{Mig2}. For atomic systems with nuclear
reactions the sudden approximation \cite{Mig1} means that the electron
density of the incident atom does not change noticeably during the nuclear
reaction in its nucleus. In addition to this, the sudden approximation for
atomic systems means that the nucleus does not move substantially during the
nuclear reaction. In general, this is true, if the nuclear reaction time
$\tau_r$ is significantly shorter than the typical atomic time $\tau_a =
\frac{\hbar}{e^4 m_e} \approx 2.418884331 \cdot 10^{-17}$ $sec$. It can be
shown that these conditions are always obeyed in the case of highly
exothermic $(n; q)-$reactions of slow neutrons with light nuclei. This means
that all changes of electron densities in arising atomic fragments can be
described in terms of the sudden approximation.

Our method is also based on the use of the sudden approximation for atomic
systems. In this work we present only a few basic equations which are used
to compute the final state probabilities for light atoms and ions. Other
details and discussion can be found in \cite{FroRui}. In general, for
few-electron atomic systems with $N-$electrons in the incident state we
compute the following integral which is called the probability amplitude:
\begin{equation}
 M_{in;fi} = \langle \Psi_{fi}({\bf x}_1, {\bf x}_2, \ldots, {\bf x}_K) \mid
 \prod_{i=1}^{K} \exp(\imath {\bf V} \cdot {\bf r}_i) \mid \Psi_{in}({\bf
 x}_1, {\bf x}_2, \ldots, {\bf x}_N) \rangle \label{equation}
\end{equation}
where $K$ is the total number of bound electrons in the final atomic system
$K \le N$. Here and below the notation ${\bf x}_i = ({\bf r}_i, \sigma_i)$
means the complete set of spatial and spin coordinates of the $i$-th
electron. If the reaction Eq.(\ref{e1}) involves the two-electron He atom
rather than the bare He nucleus, then for $K = 1$ one finds the ${}^{3}$H
atom (or ${}^{1}$H atom) in the final state, while $K = 2$ corresponds to
the two-electron ${}^{3}$H$^{-}$ ion (or ${}^{1}$H ion) in the final state.
For $K = 0$ the final system does not contain any bound electrons. The last
case (i.e. when $K = 0$) is not of interest for this study.

In the case of reaction Eq.(\ref{e1}) in the He atom (where $N = 2$) with $K
= 2$ one needs to compute the following two-electron matrix element (or
probability amplitude)
\begin{eqnarray}
 M_{in;fi} = \langle \Psi_{fi}({\bf x}_1, {\bf x}_2) \mid \exp(\imath {\bf
 V} \cdot {\bf r}_1 + \imath {\bf V} \cdot {\bf r}_2) \mid \Psi_{in}({\bf
 x}_1, {\bf x}_2) \rangle \label{e2}
\end{eqnarray}
where ${\bf V}$ is the velocity of the final nucleus formed in the reaction
Eq.(\ref{e1}). Analogously, for the reaction Eq.(\ref{e1}) in the He atom
(where $N = 2$) with $K = 1$ we need to determine a slightly different
two-electron integral (or probability amplitude):
\begin{eqnarray}
 M_{in;fi} = \langle \Psi_{fi}({\bf x}_1) \mid \exp(\imath {\bf V} \cdot
 {\bf r}_1) \mid \Psi_{in}({\bf x}_1, {\bf x}_2) \rangle \label{e3}
\end{eqnarray}
In Eqs.(\ref{e2}) and (\ref{e3}) the notation $\Psi_{in}({\bf x}_1, {\bf
x}_2)$ stands for the two-electron wave function of the helium atom, while
analogous notations $\Psi_{fi}({\bf x}_1, {\bf x}_2)$ and/or $\Psi_{fi}({\bf
x}_1)$ designate the two- and one-electron wave functions of the final
atomic systems, respectively. Below, all two-electron wave functions are
assumed to be properly antisymmetrized upon all electron variables.

Let us apply the formulas Eqs.(\ref{e2}) and (\ref{e3}) to the actual
nuclear $(n;t)-$reaction in the two-electron He atom. We shall assume that
before such a reaction the He atom was in its ground $1^1S-$state. Also, for
simplicity, in these calculations all nuclear masses are assumed to be
infinite. The main interest is to compute the ground-ground state
probabilities, since these values determine the scale of other similar
probabilities. Moreover, if the incident and final atomic systems are in
their ground states, then the corresponding probability amplitude is
relatively large. In this case it is very easy to detect and correct
possible numerical mistakes. The results of our calculations include the
corresponding probability amplitudes $M_{in;fi}$, the final state
probabilities $p_{in;fi} = \mid M_{in;fi} \mid^2$ for both nuclei formed
in the reaction Eq.(\ref{e1}), i.e. for the tritium ${}^{3}$H and protium
${}^{1}$H nuclei. Note that the velocities of these two final nuclei differ
from each other by a factor of $\sim$ 3. This explains very significant
differences between the final state probabilities determined in calculations
for the tritium and protium atoms/ions (see below).

As the starting point for numerical evaluations we can use the following
approximate expression for the ground $1^1S-$state wave function $\Psi({\bf
x}_1, {\bf x}_2)$ of the helium atom (see, e.g., \cite{March})
\begin{equation}
 \Psi({\bf x}_1, {\bf x}_2) = \frac{A^3}{\pi} \exp(-A r_1 -A r_2) (\alpha
 \beta - \beta \alpha) \label{wavef}
\end{equation}
where the $A$ value is called the effective nuclear charge. For the actual
${}^{\infty}$He atom one finds $A = Q - \frac{5}{16}$, where $Q = 2$. By
performing integration over all spin and angular variables one can produce
the following formula for the corresponding probability amplitude (for the
He $\rightarrow$ H transition during the nuclear reaction, Eq.(\ref{e1}))
\begin{equation}
 M_{in;fi} = 2 A^{\frac32} \int^{+\infty}_{0} \exp(-A r) j_{\ell}(V r)
 R_{n\ell}(\gamma r) r^2 dr \label{ampl1}
\end{equation}
where $j_{\ell}(x)$ is the Bessel function of the first kind (see, e.g.,
\cite{AS}, \cite{GR}) and $R_{n\ell}(\gamma r)$ are the radial functions of
the hydrogen-like atom/ion (see, e.g., \cite{LLQ}) formed in the final
state. If the final one-electron atomic system is in its ground state, then
we have from Eq.(\ref{ampl1})
\begin{equation}
 M_{in;fi} = 4 (A \gamma)^{\frac32} \int^{+\infty}_{0} \exp[-(A + \gamma)
 r] j_{0}(V r) r^2 dr \label{ampl2}
\end{equation}
where $V$ is the velocity of the final fragment, i.e. the velocity of the
hydrogen atom. The final state probability $P_{gg}$ is
\begin{equation}
 P_{gg} = \mid M_{in;fi} \mid^2 = \frac{64 (A \gamma)^3}{(A + \gamma)^3}
 \frac{1}{\Bigl[ 1 + \Bigl(\frac{V}{Q + q}\Bigr)^2 \Bigr]^4} \label{prob1}
\end{equation}
For the hydrogen atom the parameter $\gamma = 1$, while for the two-electron
He atom $A = \frac{27}{16}$. In this case for the tritium atom one finds
from Eq.(\ref{prob1}) $P_{gg} \approx$ 24.37809 \%. This value corresponds
to the process in which one bound $\alpha-$electron remains bound in the
final state. Since we have two bound electrons (one $\alpha-$electron and
one $\beta-$electron) in the incident He atom, then the final state
probability obtained above must be doubled, i.e. we have $P_{gg} \approx$
48.74175 \%. The use of the highly accurate wave function for the He atom
(400 exponential basis wave functions \cite{Fro98}) produces for the
$P_{gg}$ probability $\approx$ 55.19287 \%. The analogous probability for
the protium atom ${}^1$H is $\approx$ 1.02363 \%. By using the formulas
derived in \cite{FrWa09} and the approximate wave function of the He atom,
Eq.(\ref{wavef}), one can evaluate the final state probabilities for some
low-lying excited states with $n \le 3$, where $n$ is the principal quantum
number, in the tritium and protium atoms from the nuclear reaction,
Eq.(\ref{e1}). These values can be found in Table I.

It is interesting to evaluate the final state probability for the
two-electron negatively charged hydrogen ion which can be formed during
nuclear reaction, Eq.(\ref{e1}). To determine this probability we used the
highly accurate variational wave functions known for the He atom and H$^{-}$
ion \cite{Fro98}. Each of these two functions contains 400 exponents in the
relative coordinates (basis functions). The total energies produced with
such wave functions for the ground $1^1S-$states in these two systems are
-2.90372437703405 $a.u.$ for the ${}^{\infty}$He atom and -0.527751016544308
$a.u.$ for the ${}^{\infty}$H$^{-}$ ion, respectively. By performing
numerical calculations with these variational wave functions we have
evaluated the probability to form the negatively charged tritium
${}^3$H$^{-}$ ion in the nuclear $(n,{}^3$He$;t,p)-$reaction with slow
neutrons as $\approx$ 7.8680 \%. The analogous probability to form the
negatively charged protium ${}^1$H$^{-}$ ion was evaluated as $\approx$
0.06583 \%. Such a relatively large probability for the newly formed tritium
ions ${}^3$H$^{-}$ means that these ions can be detected in modern
experiments. Currently, these and other similar experiments to detect
various atomic species formed during nuclear reaction Eq.(\ref{e1}) in
two-electron helium-3 atom are critically needed to guide future theoretical
development.

\section{Bremsstrahlung from fission-type process in atomic systems}

Another group of problems is related to the analysis of radiation emitted
during the reaction of the ${}^{3}$He nuclei with slow neutrons,
Eq.(\ref{e1}). It is clear that such radiation must be similar to
radiation emitted during an arbitrary fission-type reaction in few-electron
atomic systems (atoms, ions, etc). First, let us consider a fast nuclear
fission-type reaction in a one-electron atomic system. During a
fission-type nuclear reaction the electron becomes free, and will interact
with the rapidly moving nuclear fragments. Such an interaction will produce
radiation which is, in fact, a breaking radiation or bremsstrahlung. Here we
want to derive the explicit formulas for this radiation and investigate its
spectrum. The electric charges and velocities of these fission fragments are
$Q_1 e, Q_2 e$ and $V_1, V_2$, respectively. To simplify all formulas below,
we shall assume that the both fission fragments move along the $Z-$axis.
Furthermore, in this study we restrict ourselves to the consideration of the
non-relativistic processes only. Briefly, this means that the two velocities
$V_1, V_2$ are assumed to be significantly smaller than the speed of light
$c$. The case of arbitrary velocities is discussed in \cite{Fro07}.

The second time-derivative of the dipole moment ${\bf d}$ is $\ddot{\bf d}
= e \ddot{\bf r}$, where ${\bf r} = (x,y,z)$ is the radius-vector of the
accelerated electron. In the case of a fission-type reaction in atomic
systems the explicit formula for electron's acceleration is
\begin{equation}
 \ddot{\bf r} = \frac{1}{m_e} \Bigl[ \nabla \Bigl( \frac{Q_1 e}{R_1}
 \Bigr) + \nabla \Bigl( \frac{Q_2 e}{R_2} \Bigr) \Bigr] =
 \frac{Q_1 e}{m_e} \nabla \Bigl( \frac{1}{R_1} \Bigr) +
 \frac{Q_2 e}{m_e} \nabla \Bigl( \frac{1}{R_2} \Bigr) \label{dipole}
\end{equation}
where $R_1 = \sqrt{x^2 + y^2 + (z - V_1 t)^2}$ and $R_2 = \sqrt{x^2 + y^2 +
(z + V_2 t)^2}$. The notations $V_1$ and $V_2$ stand for the velocities of
the first and second fission fragments. The second time-derivative of the
dipole moment is
\begin{equation}
 \ddot{\bf d} = \frac{Q_1 e^2}{m_e} \nabla \Bigl( \frac{1}{R_1} \Bigr) +
 \frac{Q_2 e^2}{m_e} \nabla \Bigl( \frac{1}{R_2} \Bigr) = -
 \frac{Q_1 e^2}{m_e} \frac{{\bf R_1}}{R^{3}_1} - \frac{Q_2 e^2}{m_e}
 \frac{{\bf R_2}}{R^{3}_2}
\end{equation}
where ${\bf R}_1 = (x, y, z - V_1 t)$ and ${\bf R}_2 = (x, y, z + V_2 t)$
are the three-dimensional vectors. The intensity of the non-relativistic
bremsstrahlung from a fission-type reaction in a one-electron atomic system
is
\begin{equation}
 dI = \frac{1}{4 \pi c^3} (\ddot{\bf d} \times {\bf n})^2 d\Omega =
 \Bigl( \frac{e^2}{m_e} \Bigr)^2 \frac{Q^2}{4 \pi c^3} \Bigl[
 \frac{Q_1}{Q} \frac{{\bf R_1}}{R^{3}_1} + \frac{Q_2}{Q} \frac{{\bf
 R_2}}{R^{3}_2} \Bigr]^2 sin^2\theta d\Omega
\end{equation}
where $\theta$ is the angle between the vector $\ddot{\bf d}$ and the vector
${\bf n}$ which designates the direction of propagation of radiation. The
notation $Q$ in the last equation stands for an arbitrary electric charge.
This value can be considered as an additional input parameter. In
particular, one can choose $Q = Q_1$, or $Q = Q_2$. In the case of the fast
nuclear fission-type reaction in a few-electron atomic system the last
formula takes the form
\begin{equation}
 dI = \Bigl( \frac{e^2}{m_e} \Bigr)^2 \frac{N_e Q^2}{4 \pi c^3} \Bigl[
 \frac{Q_1}{Q} \frac{{\bf R_1}}{R^{3}_1} + \frac{Q_2}{Q} \frac{{\bf
 R_2}}{R^{3}_2} \Bigr]^2 sin^2\theta d\Omega \label{e12}
\end{equation}
where $N_e$ is the number of electrons which become free after this nuclear
reaction. The formula, Eq.(\ref{e12}), can be re-written in the form
\begin{equation}
 \frac{dI}{d\Omega} = \Bigl( \frac{e^2}{m_e} \Bigr)^2 \frac{N_e Q^2}{4
 \pi c^3} \Bigl[ \frac{Q_1^2}{Q^2} \frac{1}{R^{4}_1} + \frac{Q^2_2}{Q^2}
 \frac{1}{R^{4}_2} + \frac{Q_1 Q_2}{Q^2} \Bigl(\frac{{\bf R_1} \cdot {\bf
 R_2}}{R^{3}_1 R^{3}_2}\Bigr) \Bigr]^2 sin^2\theta \label{diff}
\end{equation}
for the differential cross-section. Our derivation of these formulas for
few-electron systems is based on the fact that radiation emitted by
different post-atomic electrons is non-coherent. Indeed, different free
electrons move as independent (or `random') quantum particles. Note that
the bremsstrahlung from fast fission-type process in atomic systems can be
coherent at certain conditions, e.g., if two fission fragments move with
very large velocities, while all post-atomic electrons almost do not move at
the beginning of the process. In this case, one needs to introduce an
additional factor $N_e$ in the last formula, Eq.(\ref{diff}).

Let us discuss the spectrum of the emitted radiation. As follows from the
formula, Eq.(\ref{diff}), the intensity of bremsstrahlung from fission-type
processes in atomic systems rapidly decreases with the time $I \simeq
t^{-4}$ after the nuclear reaction which proceeds at $t = 0$. Such a
behaviour is typical for atomic systems with fast nuclear reactions. The
spectral resolution $R(\omega)$ (or spectral function) of the intensity of
dipole radiation is written in the form
\begin{equation}
 dR(\omega) = \frac{4}{3 c^3} \mid \ddot{{\bf d}}_{\omega} \mid^2
 \frac{d\omega}{2 \pi} = \frac{4 \omega^4}{3 c^3} \mid {\bf d}_{\omega}
 \mid^2 \frac{d\omega}{2 \pi} \label{spectr}
\end{equation}
where ${\bf d}_{\omega}$ is the Fourier component of the dipole moment
${\bf d}$ introduced above. As follows from Eq.(\ref{spectr}), to determine
the spectral resolution $R(\omega)$ of bremsstrahlung from a fission-type
process one needs to find the Fourier components of the dipole moment. In
this study we apply a different method and determine the Fourier components
of the $\ddot{{\bf d}}_{\omega}$ vector, Eq.(\ref{dipole}). Finally, the
problem is reduced to the calculation of the two following Fourier
transformations
\begin{equation}
 I_1(\omega; \frac{a}{V},V,\cos\eta) = \int_0^{+\infty}
 \Bigl(\frac{a^2}{V^2} \pm 2 \frac{a}{V} \cos\eta \cdot t +
 t^2 \Bigr)^{-\frac32} \exp(\imath \omega t) dt \label{ft1}
\end{equation}
and
\begin{equation}
 I_2(\omega; \frac{a}{V},V,\cos\eta) = \int_0^{+\infty}
 \Bigl(\frac{a^2}{V^2} \pm 2 \frac{a}{V} \cos\eta \cdot t + t^2
 \Bigr)^{-\frac32} t \exp(\imath \omega t) dt = -\imath
 \frac{\partial}{\partial \omega} I_1(\omega; \frac{a}{V},V,\cos\eta)
 \label{ft2}
\end{equation}
where $a$ is the `effective' radius of the original electron shell and
$\eta$ is the angle between the electron acceleration and $Z-$axis (or the
line of the nuclear motion). Note that the spectrum of the emitted radiation
depends explicitly upon $\cos\eta$. This dependence can be found for all
fast fission-type reactions and processes. The lower limit in these formulas
is zero, but, in reality, we cannot use times $t$ which are shorter than
$\tau = \frac{a}{c}$. Formally, this means that the lower limits in
Eqs.(\ref{ft1}) and (\ref{ft2}) can slightly be changed and this can be
used to simplify the explicit formulas for the corresponding Fourier
components. The upper limits in Eqs.(\ref{ft1}) and (\ref{ft2}) can also be
chosen to be finite. For instance, it is possible to obtain a very good
approximation for the integrals, Eqs.(\ref{ft1}) and (\ref{ft2}), by using
the upper limit $\frac{10 a}{V}$. In actual situations one can use the
formula $\exp(\imath x) = \cos x + \imath \sin x$ and then calculate all
arising integrals numerically, or by using analytical formulas from the
Tables of Fourier $sine/cosine$ transformations (see, e.g., \cite{Transf}).
A good analytical approximation for the integral, Eq.(\ref{ft1}), is
\begin{eqnarray}
 I_1(\omega; \frac{a}{V},V,\cos\eta) \approx \exp\Bigl(-\imath \omega
 \frac{a}{V} \cos\eta\Bigr) \int_{0}^{+\infty} [ b^{2} + \tau^2 ]^{-\frac32}
 \exp(\imath \omega \tau) d\tau \nonumber \\
 = \exp\Bigl(-\imath \omega \frac{a}{V} \cos\eta\Bigr)
 \Bigl[\frac{\omega}{b} K_1(b \omega) + \frac{1}{\omega} \ln\Bigl(\frac{1
 + \frac12 b \omega}{1 - \frac12 b \omega}\Bigr) - \frac{\pi}{4}
 \frac{\omega^2}{\Bigl(1 - \frac12 b \omega\Bigr)^2} \Bigr] \label{radiat}
\end{eqnarray}
where $b = \frac{a \sin\eta}{V}$ and $K_1(x)$ is the MacDonald function
(see, e.g., \cite{Wats}). The derivative of Eq.(\ref{radiat}) with respect
to the frequency $\omega$ allows one to determine the $I_2(\omega;
\frac{a}{V},V,\cos\eta)$ integral, Eq.(\ref{ft2}), and obtain the analytical
formula for the spectral resolution $R(\omega)$ of the intensity of
bremsstrahlung, Eq.(\ref{spectr}), from the fast fission-type (fast) process
in few- and many-electron atomic systems. The derivation of explicit
expression for the spectral resolution $R(\omega)$ is straightforward, but
the final formula is cumbersome and it is not given here. Note only that the
term $\sim Q_1 Q_2$ in this formula represents interference of radiation
which is emitted by the two rapidly moving fission fragments. In application
to the nuclear fission of heavy nuclei, e.g., ${}^{239}$Pu, ${}^{247}$Cm,
etc, the spectral resolution $R(\omega)$ must be averaged over the actual
mass/charge distribution known for each fissionable nucleus.

\section{Conclusion}

We have considered the nuclear reaction, Eq.(\ref{e1}), involving the
two-electron ${}^3$He atom and slow neutrons. To determine the final state
probabilities for one- and two-electron tritium and protium atoms/ions we
have developed a new method which is based on direct computation of
probability amplitudes. This method is much simpler than an alternative
procedure based on the use of few-electron density matrices. Furthermore, it
allows one to determine the final state probabilities for various bound
states in the incident and final atomic systems. By using this method we
have found that the probability to form the tritium atom ${}^3$H in its
ground state is $\approx$ 55.19287 \%, while analogous probability to form
the protium atom in its ground state ${}^1$H is $\approx$ 1.02363 \%.
Analogous probabilities for the negatively charged hydrogen ions, i.e. for
the ${}^3$H$^{-}$ and ${}^1$H$^{-}$ ions to be formed in the nuclear
$(n,{}^3$He$; t,p)-$reaction with slow neutrons, are $\approx$ 7.8680 \% and
$\approx$ 0.06583 \%, respectively. The corresponding uncertainties for
such probabilities can be evaluated as $\approx 0.3 - 1 \cdot 10^{-4}$ \%
for the ground states and as $\approx 0.5 - 1 \cdot 10^{-3}$ \% for the
excited states. Another important problem is the analysis of bremsstrahlung
from fast fission-type reactions in atomic systems. In this study such an
analysis is performed for the non-relativistic case. Its generalization to
the systems with relativistic velocities is quite complicated (see, e.g.,
\cite{Fro07}).

In the future it is necessary to study analogous nuclear reactions involving
atoms with the ${}^{6}$Li, ${}^{10}$B, ${}^{14}$N (and some other) nuclei.
For such atomic systems we need to determine the final state probabilities
to form various few-electron atoms and ions in the reaction of these atoms
with slow neutrons. These results are of interest in numerous applications
of these nuclear reactions, e.g., in the Boron Neutron Capture Therapy
(BNCT) where the reaction ($n, {}^{10}$B; ${}^{7}$Li,${}^{4}$He) is used
\cite{Haw1} - \cite{Haw3}. Nuclear reaction Eq.(\ref{e1}) and the analogous
reaction with the ${}^{6}$Li nuclei are extensively used in thermonuclear
explosive devices \cite{FrWa09}. Also, these nuclear reactions are used in
various detectors (see, e.g., \cite{WW}, \cite{Yamp}, \cite{Tsoul}) of
thermal and slow neutrons. By using our results for the final state
probabilities of different atomic species formed in such reactions we can
improve the overall sensitivity of such detectors.

\begin{center}
    {\bf Acknowledgements}
\end{center}

It is a pleasure to thank Professor M. Frederick Hawthorne for useful
references and the University of Western Ontario for financial support.

\newpage
  \begin{table}[tbp]
   \caption{The probabilities (in \%) of the final states in the
            tritium/protium atom arising in the exothermic nuclear
            reaction, Eq.(1), of the two-electron helium-3 atom with
            slow neutrons. Here the notation $a(b)$ stands for $a
            \cdot 10^{-b}$.}
     \begin{center}
     \begin{tabular}{lllllll}
        \hline\hline
 atom/state & $1s$   &  $2s$    &  $2p$    &  $3s$    &  $3p$    &  $3d$ \\
           \hline
 ${}^{3}$H & 48.7418 & 9.42077  & 0.803701 & 0.534751 & 0.053371 & 0.032033 \\

 ${}^{1}$H & 0.53715 & 0.105205 & 0.002480 & 0.008709 & 1.28725$\cdot 10^{-2}$ & 3.7800$\cdot 10^{-6}$ \\
        \hline\hline
  \end{tabular}
  \end{center}
  \end{table}
\end{document}